\RequirePackage{fix-cm}
\RequirePackage{graphicx}
\documentclass[preprint,12pt]{elsarticle}

\usepackage{float}
\usepackage{makecell}
\usepackage{tabularx}
\usepackage{hhline}
\usepackage{colortbl}
\usepackage{array}
\usepackage[dvipsnames,table]{xcolor}
\definecolor{pastelred}{rgb}{1.0, 0.6, 0.6}
\definecolor{pastelgreen}{rgb}{0.6, 1.0, 0.6}
\definecolor{pastelorange}{rgb}{1.0, 0.8, 0.6}
\definecolor{pastelyellow}{rgb}{1.0, 1.0, 0.8}
\usepackage[colorlinks=true,linkcolor=blue,citecolor=blue,urlcolor=blue]{hyperref}
\usepackage[separate-uncertainty,retain-explicit-plus,per-mode=symbol,range-phrase=\mbox{--},range-units=single]{siunitx}
\usepackage[export]{adjustbox}
\usepackage{wrapfig}
\usepackage{ragged2e}
\usepackage{array,mathtools,dcolumn}
\usepackage{amsmath,amssymb}
\usepackage[below]{placeins}
\usepackage[table]{xcolor}
\usepackage{soul}
\usepackage{tikz}
\usepackage{afterpage}
\usepackage{lineno}
\usepackage{listings}
\usepackage{array}
\usepackage[version=4]{mhchem}
\usepackage{multirow}
\usepackage{eurosym}
\usepackage{pagecolor}
\usepackage{fancyhdr}
\usepackage{etoolbox}
\usepackage{calc}
\usepackage{dcolumn}

\usepackage{bm}
\usepackage{xargs}
\usepackage{xfrac}
\newcommandx{\unsure}[2][1=]{\todo[linecolor=red,backgroundcolor=red!25,bordercolor=red,#1]{#2}}
\newcommandx{\change}[2][1=]{\todo[linecolor=blue,backgroundcolor=blue!25,bordercolor=blue,#1]{#2}}
\newcommandx{\info}[2][1=]{\todo[linecolor=OliveGreen,backgroundcolor=OliveGreen!25,bordercolor=OliveGreen,#1]{#2}}
\newcommandx{\improvement}[2][1=]{\todo[linecolor=Plum,backgroundcolor=Plum!25,bordercolor=Plum,#1]{#2}}
\newcommandx{\thiswillnotshow}[2][1=]{\todo[disable,#1]{#2}}

\usepackage{paralist}
\usepackage{tcolorbox}
\usepackage{pgfornament}
\usepackage{appendix}
\usepackage{enumitem}
\setdescription{itemsep=0pt,parsep=0pt,labelsep=.2cm,leftmargin=2.2cm,labelwidth=2cm,listparindent=.7cm}
\setlist{nolistsep,leftmargin=1cm}
\newlist{enumcompactitem}{itemize}{3}
\setlist[enumcompactitem]{topsep=0pt,partopsep=0pt,itemsep=0pt,parsep=0pt}
\setlist[enumcompactitem,1]{label=\textbullet}
\setlist[enumcompactitem,2]{label=---}
\setlist[enumcompactitem,3]{label=*}
\newlist{enumcompactdesc}{description}{3}
\setlist[enumcompactdesc]{topsep=0pt,partopsep=0pt,itemsep=0pt,parsep=0pt}
\newlist{enumcompactenum}{enumerate}{3}
\setlist[enumcompactenum]{topsep=0pt,partopsep=0pt,itemsep=0pt,parsep=0pt}
\setlist[enumcompactenum,1]{label=\arabic*}
\setlist[enumcompactenum,2]{label=\alph*}
\setlist[enumcompactenum,3]{label=\roman*}
\newcommand{\Geant}{\mbox{\tt Geant4}}

\newcommand{\GFNDL}{\mbox{\tt G4NDL}}
\newcommand{\GFPE}{\mbox{\tt G4PhotonEvaporation}}

\newcommand{\CASCADE}{\mbox{\tt G4CASCADE}}
\newcommand{\NuDEX}{\mbox{\tt NuDEX}}

\newcommand{\ENSDF}{\mbox{\tt ENSDF}}
\newcommand{\CapGam}{\mbox{\tt CapGam}}
\newcommand{\XUNDL}{\mbox{\tt XUNDL}}
\newcommand{\RIPL}{\mbox{\tt RIPL-3}}
\newcommand{\DICEBOX}{\mbox{\tt DICEBOX}}

\DeclareSIUnit\c{\mbox{$c$}}
\DeclareSIUnit\magn{\mbox{$\times$}}
\DeclareSIUnit\min{min}
\DeclareSIUnit\week{week}
\DeclareSIUnit\month{mo}
\DeclareSIUnit\months{mos}
\DeclareSIUnit\year{yr}
\DeclareSIUnit\years{years}
\DeclareSIUnit\yr{yr}
\DeclareSIUnit\standard{std}
\DeclareSIUnit\str{sr}
\DeclareSIUnit\ppm{ppm}
\DeclareSIUnit\ppb{ppb}
\DeclareSIUnit\ppt{ppt}
\DeclareSIUnit\pe{PE}
\DeclareSIUnit\spe{SPE}
\DeclareSIUnit\pdm{PDM}
\DeclareSIUnit\ev{events}
\DeclareSIUnit\ct{counts}
\DeclareSIUnit\neutron{\mbox{$n$}}
\DeclareSIUnit\smp{samples}
\DeclareSIUnit\Sample{S}
\DeclareSIUnit\ch{ch}
\DeclareSIUnit\hit{hit}
\DeclareSIUnit\hits{hits}
\DeclareSIUnit\bin{(\mbox{5-PE}~bin)}
\DeclareSIUnit\sgm{\mbox{$\sigma$}}
\DeclareSIUnit\rms{RMS}
\DeclareSIUnit\keVee{\mbox{keV$_{e{\rm e}}$}}
\DeclareSIUnit\keVr{\mbox{keV$_{\rm nr}$}}
\DeclareSIUnit\eVee{\mbox{eV$_{\rm ee}$}}
\DeclareSIUnit\eVr{\mbox{eV$_{\rm nr}$}}
\DeclareSIUnit\ph{photon}
\DeclareSIUnit\el{\mbox{$e^-$}}
\DeclareSIUnit\pm{\mbox{PMT}}
\DeclareSIUnit\pixel{\mbox{pixel}}
\DeclareSIUnit\inch{''}
\DeclareSIUnit\foot{'}
\DeclareSIUnit\bit{bit}
\DeclareSIUnit\sample{samples}
\DeclareSIUnit\barn{barn}
\DeclareSIUnit\bara{bar}
\DeclareSIUnit\barg{barg}
\DeclareSIUnit\mlardepth{\mbox(meter~of~\LAr~depth)}
\DeclareSIUnit\Curie{Ci}
\DeclareSIUnit\psi{psi}
\DeclareSIUnit\psf{psf}
\DeclareSIUnit\pcf{pcf}
\DeclareSIUnit\parsec{pc}
\DeclareSIUnit\mwe{\mbox{m.w.e.}}
\DeclareSIUnit\liveday{\mbox{live-days}}
\DeclareSIUnit\days{\mbox{days}}
\DeclareSIUnit\miles{\mbox{miles}}
\DeclareSIUnit\lumens{\mbox{lm}}
\DeclareSIUnit\degreeC{\mbox{$^{\circ}$C}}
\DeclareSIUnit\degreeF{\mbox{$^{\circ}$F}}
\DeclareSIUnit\electron{\mbox{$e^-$}}
\DeclareSIUnit\Euro{\mbox{\euro}}
\DeclareSIUnit\cph{cph}
\DeclareSIUnit\neq{neq}
\DeclareSIUnit\normal{\mbox{N}}

\newcommand{\ngamma}{\mbox{($n$,$\gamma$)}}

\newcommand{\bdecay}{\mbox{$\beta$-decay}}

\newcommand{\znbb}{\mbox{$0\nu\beta\beta$}}

\newcommand{\DSk}{\mbox{DarkSide-20k}}

\newcommand{\DEAP}{\mbox{DEAP-3600}}

\newcommand{\SNOp}{\mbox{SNO+}}

\newcommand{\SiPM}{\mbox{SiPM}}

\newcommand{\LAr}{\ce{LAr}}

\newcommand{\LXe}{\ce{LXe}}

\newcommand{\gr}{\mbox{$\gamma$-ray}}
\newcommand{\grs}{\mbox{$\gamma$-rays}}

\newcommand{\xrs}{\mbox{X-rays}}

\journal{Nuclear Instruments and Methods A}
\definecolor{myRed}{HTML}{CC0000}
\definecolor{myBlue}{HTML}{0088EE}
\definecolor{myPurple}{HTML}{6600CC}

\begin{document}

\title{G4CASCADE: A data-driven implementation of \ngamma\ cascades in Geant4}

\author[UCR,WVU]{Leo Weimer}
\author[Queens]{Emma Ellingwood}
\author[Claremont]{Otis Fischer}
\author[UCR]{Michela Lai}
\author[UCR]{Shawn Westerdale}
\affiliation[UCR]{organization={Department of Physics \& Astronomy, University of California, Riverside}, 
           city={Riverside},
          postcode={92521},
         state={CA},
          country={USA}}
\affiliation[WVU]{organization={Department of Physics \& Astronomy, West Virginia University}, 
           city={Morgantown},
          postcode={26506},
         state={West Virginia},
          country={USA}}

\affiliation[Queens]{organization={Department of Physics, Engineering Physics and Astronomy, Queen's University}, 
           city={Kingston},
          postcode={K7L~3N6},
         state={Ontario},
          country={Canada}}

\affiliation[Claremont]{organization={W.~M. Keck Science Department, Claremont McKenna College}, 
           city={Claremont},
          postcode={91711},
         state={California},
          country={USA}}
\date{\today}

\begin{abstract}
De-excitation \gr\ cascades from neutron captures form a dominant background to MeV-scale signals. 
The \Geant\ Monte Carlo simulation toolkit is widely used to model backgrounds in nuclear and particle physics experiments.
While its current modules for simulating \ngamma\ signals, \GFNDL\ and \GFPE, are excellent for many applications, they do not reproduce known \gr\ lines and correlations relevant at \SIrange{2}{15}{\MeV}. 
\CASCADE\ is a new data-driven \Geant\ module that simulates  \ngamma\ de-excitation pathways, with options for how to handle shortcomings in nuclear data. 
Benchmark comparisons to measured \gr\ lines and level structures in the \ENSDF\ database and other sources show significant improvements, with decreased residuals and full energy conservation.
This manuscript describes the underlying calculations performed by \CASCADE, its various usage options, and benchmark comparisons.
\CASCADE\ for \Geant\texttt{-10} is available on GitHub at\\\url{https://github.com/UCRDarkMatter/CASCADE}


\end{abstract}

\maketitle


\section{Introduction}
\label{Sec:introduction}

Neutrons that capture on nuclei produce a de-excitation \gr\ cascade that is important to particle and nuclear physics experiments.
Antineutrino detectors that use inverse \bdecay\ often rely on delayed coincidences between positron and \ngamma\ signals from the outgoing neutron capturing, often on \ce{^1H}, \ce{^12C},  \ce{^16O}, and \ce{^157Gd}, among others.
Neutrons from detector materials and the environment produce the dominant \SIrange{3}{11}{\MeV} backgrounds for signals including some neutrino-nucleus interactions and neutrinoless double \bdecay\ (\znbb)~\cite{Tornow:2015rqa}. 

\Geant\ is a widely-used toolkit for simulating particle interactions with matter~\cite{Agostinelli:2002hh}.
While very successful, it has some issues simulating \ngamma\ signals, largely due to limitations in nuclear data and models.
With current models, users therefore choose between reproducing measured \gr\ lines and conserving the total energy of the de-excitation cascade, and inaccuracies often arise in both due to incomplete knowledge of \gr\ correlations.

\CASCADE\ (the \Geant\ Code for Allowing Simulation of n-Capture and De-excitation with \ENSDF) is a new data-driven module that can integrate with existing \Geant\ neutron transport models to implement the full de-excitation pathways for \ngamma\ reactions~\cite{UCRDarkMatterCASCADERepository}.
\CASCADE\ uses data in the Evaluated Nuclear Structure Data File (\ENSDF) evaluations~\cite{ensdf:2023} and recent measurements to simulate \gr\ cascades. 
In Section~\ref{Sec:Geant4}, current \ngamma\ models in \Geant\ are summarized and their limitations are discussed. 
Section~\ref{Sec:software} describes the implementation of \CASCADE.
Example \gr\ spectra and benchmarks are illustrated in Section~\ref{Sec:results}, focusing on representative isotopes important to low-background experiments.
Lastly, Section~\ref{Sec:discussion} discusses these results and future developments planned for the code.
\CASCADE\ is currently available on GitHub at
\mbox{\url{https://github.com/UCRDarkMatter/CASCADE}}.

\section{Current models in \Geant}
\label{Sec:Geant4}

\Geant\ provides the user with two modules for simulating radiative neutron captures: \GFNDL, which references \gr\ intensities from a nuclear data library, and \GFPE, which generates the cascade from the capture level using statistical models.

\begin{figure}[htb]
    \centering
    \includegraphics[width=\linewidth]{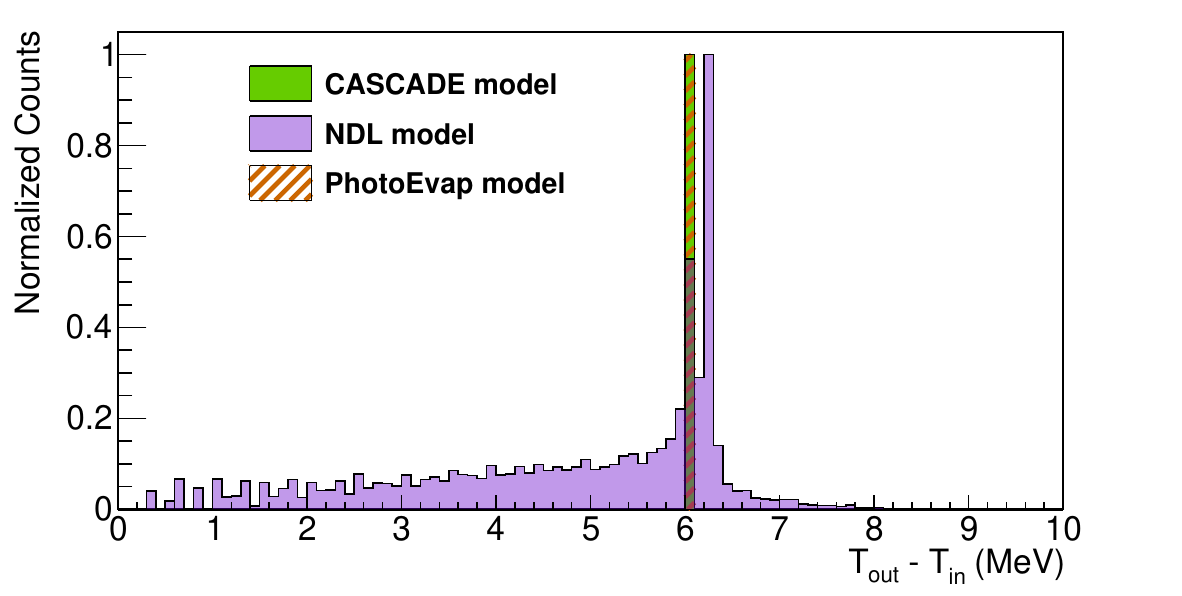}
    \caption{Kinetic energy difference between incident and outgoing particles for each model for \ce{^40Ar}\ngamma.}
    \label{fig:gd_ndl}
\end{figure}

\GFNDL\ models \SI{<20}{\MeV} neutron transport using data libraries for cross-sections, probability functions, and energy-angle distributions for emitted particles, where available. 
Fig.~\ref{fig:gd_ndl} shows the total kinetic energy of outgoing particles in excess of the incident particles' energy for $^{40}$Ar\ngamma, simulated using \GFNDL, \GFPE, and \CASCADE. 
While \GFPE\ and \CASCADE\ produce nearly mono-energetic peaks at the reaction's Q-value, as expected, \GFNDL\ produces a broad distribution, illustrating its violation of energy conservation on event-by-event bases.
\GFPE\ conserves energy but often has extra and missing \grs\ or incorrect intensities, as seen in Fig.~\ref{fig:argon_cascade}.

\section{The \CASCADE\ software}
\label{Sec:software}

Daughter nuclei energy levels and transitions, including branching ratios and conversion coefficients, are input to \CASCADE\ from an \ENSDF-derived library.
\CASCADE\ integrates into \Geant's high precision neutron modules (\texttt{G4NeutronHP}), and runs if data exist for a given isotope and if a user-defined environment variable is set to true. Otherwise, \GFNDL\ and/or \GFPE\ are used as normal.

After a neutron is captured, \CASCADE\ begins the de-excitation cascade at the energy level nearest the neutron separation energy. Any excess energy is dissipated by a \gr\ or is ignored, depending on user preference.
\CASCADE\  simulates nuclear de-excitation by moving down the energy levels randomly, weighted by the branching ratios for each transition. 
For each transition, \grs\ are released with energy equal to the difference between the initial and final levels. 
Doing so allows for the correlated emission of \grs\ with appropriate intensities for many isotopes. 
Level structures were generated by converting \ENSDF\ thermal neutron capture data for each available isotope. 
A binary serialization format is used for this database for faster loading times. 

Branching ratios are calculated based on several factors in the \ENSDF\ files. 
For each transition, the Relative Intensity ($\texttt{RI}$) and the Total Intensity ($\mathtt{TI}$) are multiplied by normalization factors, $\mathtt{NR}$ and $\mathtt{NT}$. 
If $\texttt{RI}$ is defined, the branching ratio BR is
\begin{gather}
  \begin{aligned}
     \text{BR} &= \mathtt{RI}.
  \end{aligned}
\end{gather}
If $\texttt{RI}$ is not defined but $\mathtt{TI}$ is, then
\begin{gather}
  \begin{aligned}
     \text{BR} &= \frac{\mathtt{TI}}{1+\mathtt{CC}}.
  \end{aligned}
  \label{eq:br}
\end{gather}
If the conversion coefficient ($\mathtt{CC}$) is defined, two transitions to the same level are possible, with branching ratios for \gr\ and conversion electron emission, $\text{BR}_\gamma$ and $\text{BR}_e$, given by, 
\begin{gather}
  \begin{aligned}
     \text{BR}_\gamma &= \mathtt{RI} \text{ or } \frac{\mathtt{TI}}{1+\mathtt{CC}},\\
     \text{BR}_e &= \text{BR}_\gamma \times \mathtt{CC} .
  \end{aligned}
  \label{eq:br_cc}
\end{gather}

If an inner shell conversion electron is emitted, the atomic relaxation and Auger cascades are simulated using \texttt{G4RDAtomicDeexcitation}.
When both $\mathtt{RI}$ and $\mathtt{TI}$ are blank, the transition is removed.

\ENSDF\ lists some \grs\ that cannot be placed in the level structure. 
\CASCADE\ ignores them by default, but can optionally add them at the top energy level. 
Unplaced \gr\ and conversion electron transitions' branching ratios are calculated as in Equations~\eqref{eq:br} and~\eqref{eq:br_cc}. 
Since these transitions end at an energy level not in the level structure, \GFPE\ is used to complete the de-excitation cascade to the ground state.
This option is referred to as \CASCADE\ (PE).

After capturing a fast neutron, a nucleus may have energy above the highest known energy level.
Since unbound levels are unknown for most nuclei, approximations must be made.
\CASCADE\ provides users with two options. 
First, the energy of the nucleus can be set to the highest available energy level within the level structure, ignoring the excess energy.
This results in a small violation of energy conservation, typically \SI{<200}{\eV}, although this may be larger if the neutron has more kinetic energy at the time it is captured.
Alternatively, the excess energy may be dissipated as a single \gr; doing so conserves energy, though introduces \gr\ lines that have not been experimentally observed. 
  
\section{Benchmark comparisons}
\label{Sec:results}
To benchmark \CASCADE's performance, \gr\ intensities are compared to those cataloged in the \CapGam\ database~\cite{pritychenko2006nuclear} (which uses \ENSDF\ and the eXperimental Unevaluated Nuclear Data List, \XUNDL~\cite{xundl}) for isotopes relevant to rare event searches, alongside calculation with \GFNDL, \GFPE, and \NuDEX\ \cite{nudex}, an \ngamma\ simulation tool (with no \Geant\ integration) based on \ENSDF\ and nuclear structure data from \RIPL, supplemented with the \DICEBOX\ statistical nuclear model calculations.
For each model, \gr\ spectra were obtained by simulating \SI{1}{\eV} neutrons capturing on each isotope using \texttt{Geant4-10.7.4}.
Benchmarks are chosen to reflect key reactions and to show the range of \CASCADE's performance, including isotopes with the most and least agreement with \CapGam. 

\begin{figure}[htb]
    \centering
    \includegraphics[width=\linewidth]{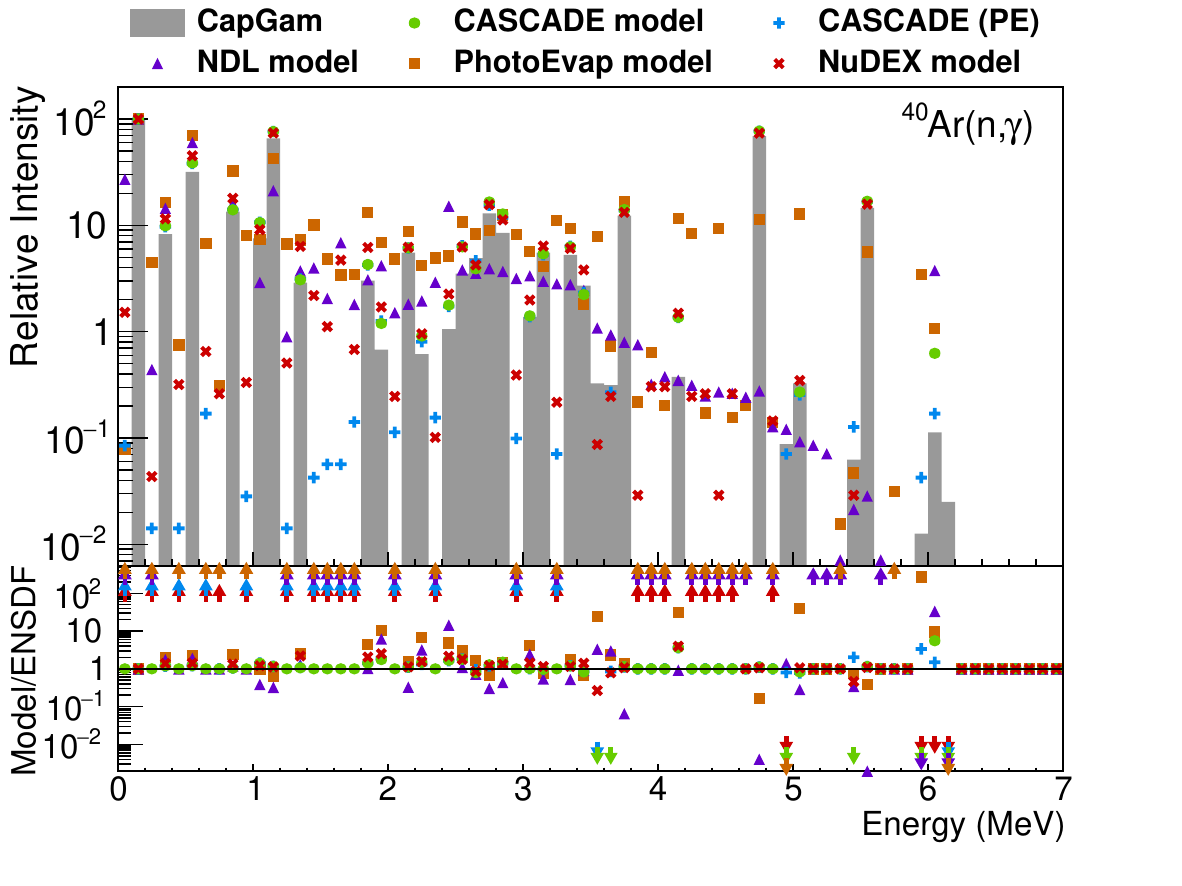}
    \caption{Comparison of \gr\ intensities for \ce{^40Ar}\ngamma\ in \CapGam, \CASCADE, and other models. }
    \label{fig:argon_cascade}
\end{figure}

\begin{figure}[h!]
    \centering
    \includegraphics[width=0.9\linewidth]
    {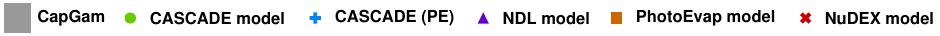}
    \includegraphics[width=0.49\linewidth]{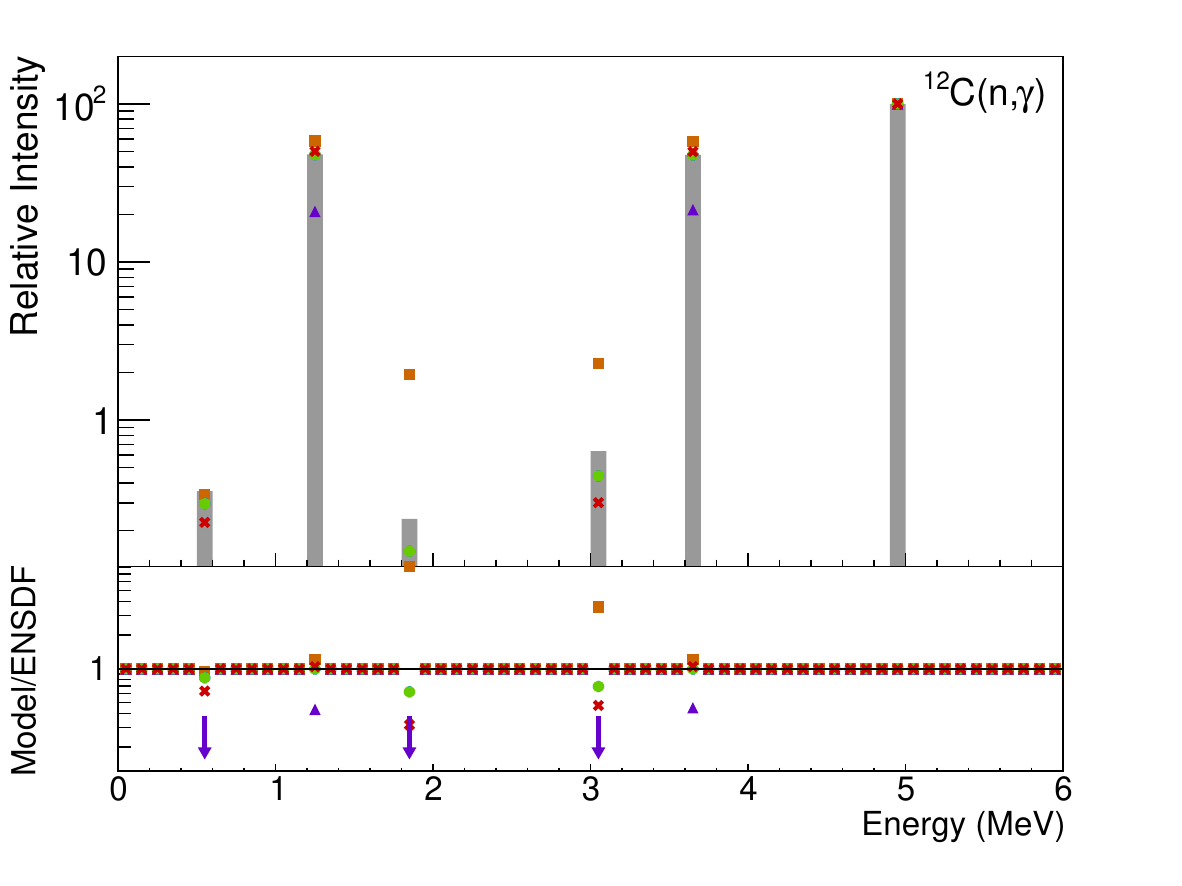}
    \includegraphics[width=0.49\linewidth]{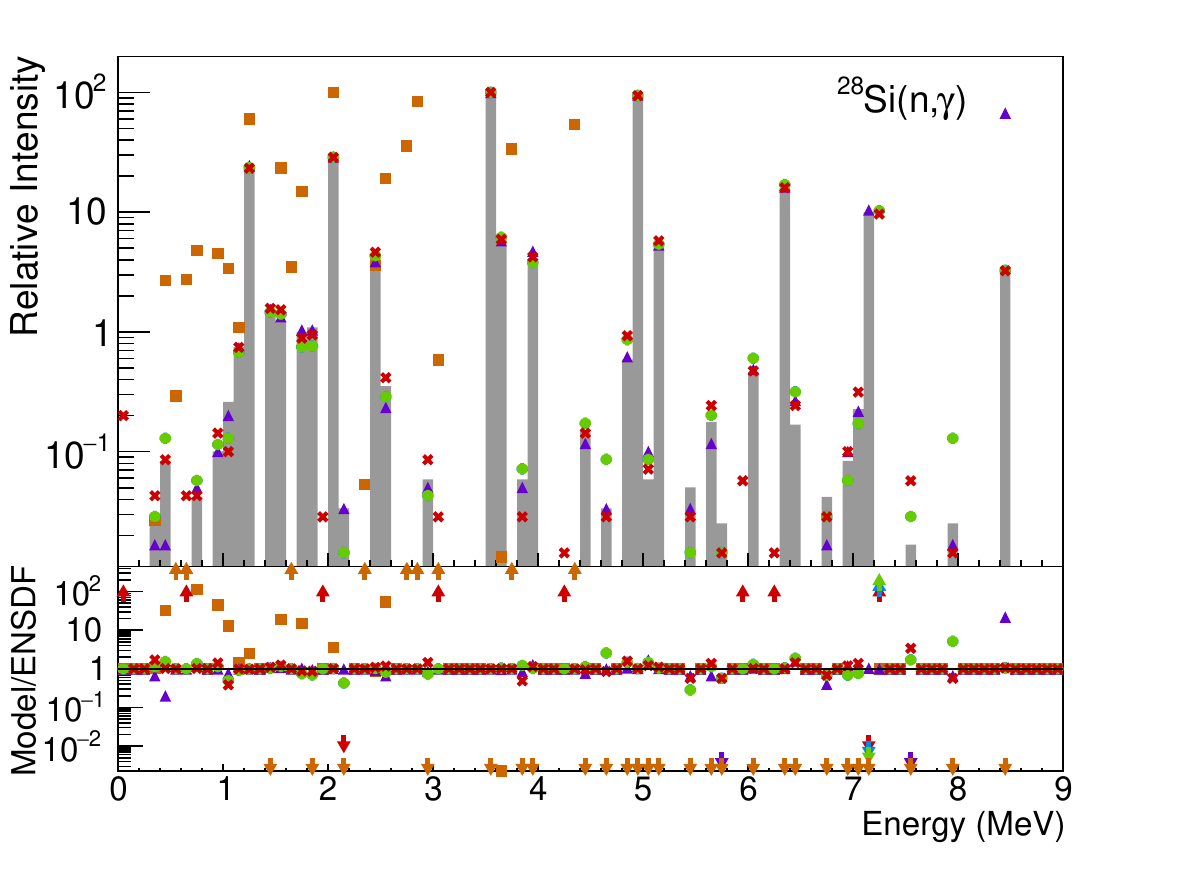}
    \includegraphics[width=0.49\linewidth]{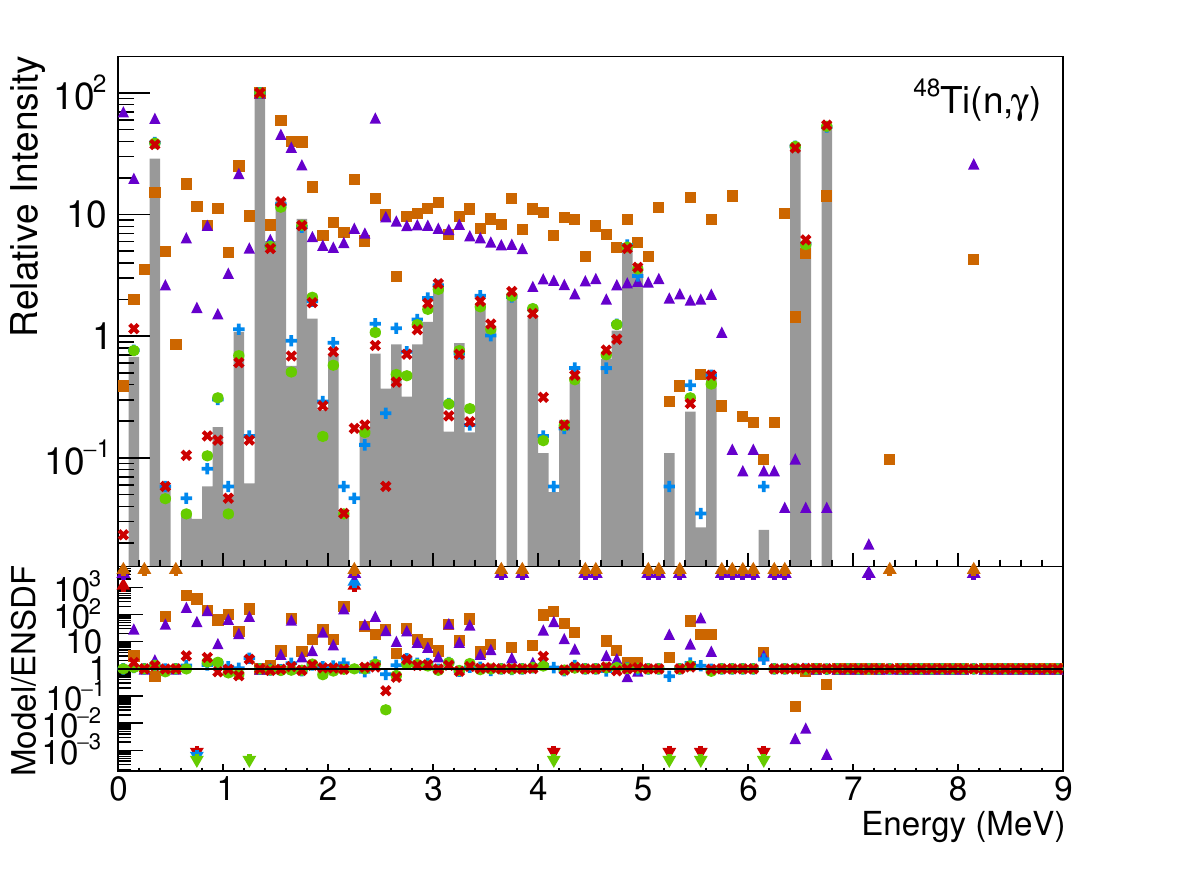}
    \includegraphics[width=0.49\linewidth]{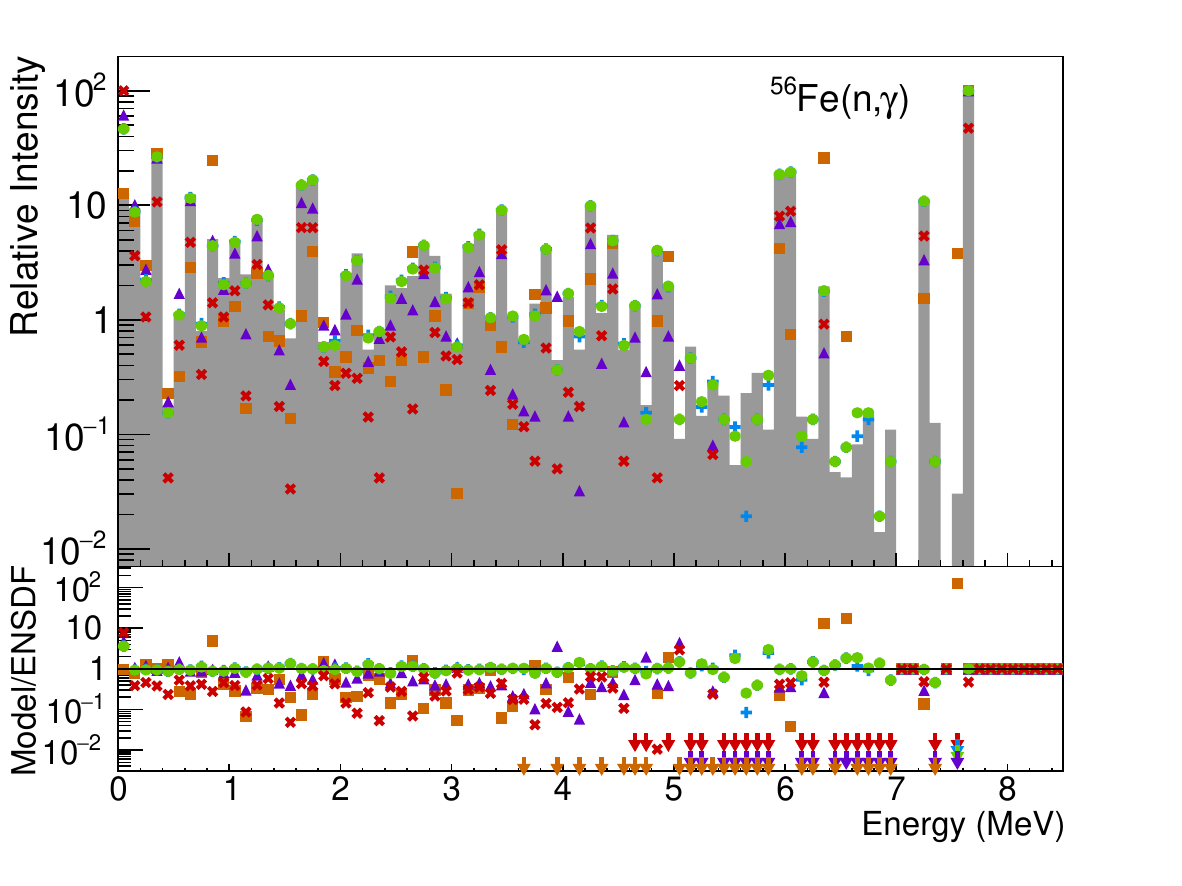}
    \includegraphics[width=0.49\linewidth]{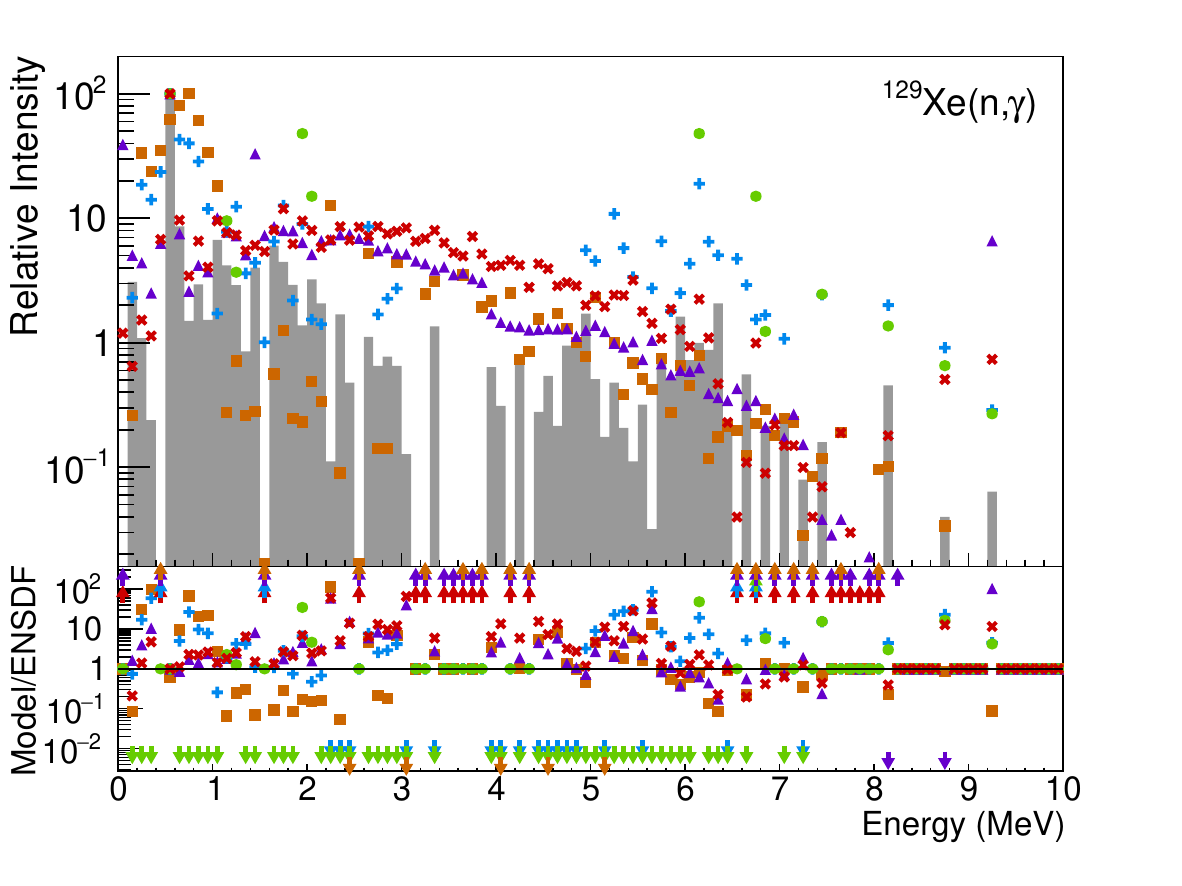}
    \includegraphics[width=0.49\linewidth]{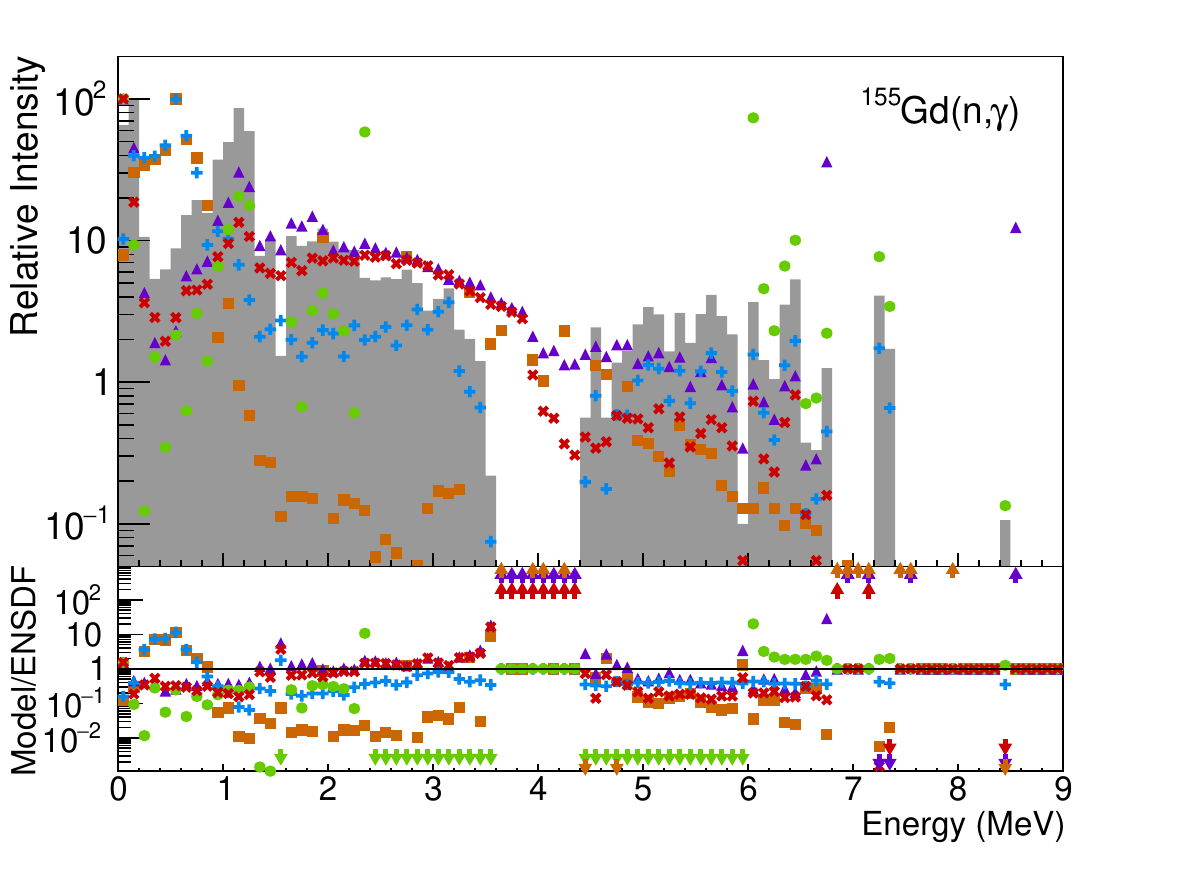}
    \caption{Data/model comparisons: (top left) \ce{^12C}, (top right) \ce{^28Si}, (mid left) \ce{^48Ti}, (mid right) \ce{^56Fe}, (bottom left) \ce{^129Xe}, (bottom right) \ce{^155Gd}. ``CASCADE (PE)'' uses \CASCADE\ with \GFPE.}
    \label{fig:xenon_cascade}
    \label{fig:iron_cascade}
    \label{fig:titanium_cascade}
    \label{fig:silicium_cascade}
    \label{fig:carbonium_cascade}
    \label{fig:gadolinium_cascade}
\end{figure}

Liquid argon (\LAr) and xenon (\LXe) are popular targets for dark matter and neutrino detectors, such as the \LAr-based dark matter direct detection experiments DarkSide-50~\cite{darksidecollaborationDarkSide50532dayDark2018a}, DEAP-3600~\cite{deapcollaborationSearchDarkMatter2019}, and DarkSide-20k~\cite{aalsethDarkSide20k20Tonne2018}, as well as the \LAr\ neutrino detectors ICARUS \cite{ICARUS:2001vle}, LAr1-ND~\cite{LArTPC:2013hbn}, MicroBooNE~\cite{acciarriProposalThreeDetector2015}, and the DUNE far detectors~\cite{capozziDUNENextGenerationSolar2019}.
\LXe\ detectors are widely used for dark matter direct detection experiments XENONnT \cite{xenoncollaborationFirstDarkMatter2023}, LZ~\cite{aalbersFirstDarkMatter2023}, PandaX-4T~\cite{mengDarkMatterSearch2021}, and XMASS-I~\cite{xmasscollaborationDirectDarkMatter2023}, and for \znbb\ searches like nEXO~\cite{nexocollaborationSensitivityDiscoveryPotential2018}.
Figures~\ref{fig:argon_cascade} and~\ref{fig:xenon_cascade} (bottom left) show \gr\ intensities for \ce{^40Ar}\ngamma\ and \ce{^129Xe}\ngamma\ from \CapGam\ compared with those generated by \CASCADE\ (with and without unplaced gammas), \GFNDL, \GFPE, and \NuDEX.
\CASCADE\ reproduces \ce{^40Ar}\ngamma\ measurements better than \GFPE, \GFNDL, and \NuDEX, and reproduces \ce{^129Xe}\ngamma\ measurements better than \GFPE, while conserving energy and correlating \grs\ for both.

Similar results are shown also for \ce{^56Fe} and \ce{^48Ti} in Fig.~\ref{fig:iron_cascade} (middle right and bottom left). 
These isotopes are abundant in structural components, such as cryostats used in \LAr\ and \LXe\ detectors, as well as for  planned \znbb\ search in \ce{ZnO}-based scintillating bolometers~\cite{56Fe_bolometers}, and in \ce{Ge}-based detectors used for \znbb\ searches like LEGEND~\cite{abgrallLargeEnrichedGermanium2017}, in addition to reactor neutrino measurements like RICOCHET~\cite{56Fe_Cevns}, and other neutrino experiments~\cite{56Fe_snowmass}.

Silicon photomultiplier (\SiPM) photosensors are increasingly used in particle detectors, with foreseen uses in DUNE~\cite{silicium_DUNE}, DarkSide-20k~\cite{dinceccoDevelopmentNovelSingleChannel2018}, and Time-of-Flight Positron Emission Tomography (TOF-PET) medical imaging scanners~\cite{siPM_petalo,sipm_3Dpi,siPm_mephi}. 
The SuperCDMS dark matter direct detection experiment also plans to use a silicon target~\cite{albakryFirstMeasurementNuclearrecoil2023}.
As shown in Fig~\ref{fig:silicium_cascade} (middle left), \CASCADE's data-driven approach offers the most accurate relative intensity spectrum, comparable with results of \NuDEX.

Acrylic and PTFE are also common structural materials. 
They are used in the \znbb\ search with \ce{^136Xe} in KamLand-Zen~\cite{acrylic_KamLAND-Zen}, and they contain \LAr\ in \DEAP\ and \DSk.
Since acrylic and PTFE are rich in carbon, as are organic liquid scintillators used in detectors like the \SNOp\ \znbb\ experiment~\cite{collaborationSNOExperiment2021}, \ce{^12C}\ngamma\ is an important reaction.
Due to its high \ngamma\ cross section, gadolinium is also commonly used to detect neutrons.
The \DSk\ Collaboration developed \ce{Gd}-loaded acrylic for vetoing neutron backgrounds~\cite{gadolinium_ds20k,collaborationNewHybridGadolinium2024}, and the Super-Kamiokande experiment will use \ce{Gd}-loaded water to detect inverse \bdecay~\cite{simpsonSensitivitySuperKamiokandeGadolinium2019}.
Figs.~\ref{fig:carbonium_cascade} (top left and bottom right) benchmark \CASCADE's de-excitation cascade simulations for captures on \ce{^12C} and \ce{^155Gd}.

\renewcommand{\arraystretch}{1.5}
\begin{table*}[htbp]
   \centering
   \small
   \arrayrulecolor{black} 
    \caption{Summary comparisons showing how all five models compare with \gr\ lines in \CapGam, color-coded according to RMS (\textcolor{myBlue}{blue} for lowest, \textcolor{myBlue!50!myPurple}{indigo} second, \textcolor{myPurple}{purple} third, 
    \textcolor{myRed!50!myPurple}{magenta} fourth, and \textcolor{myRed}{red} for highest). Three metrics summarize residuals ($\text{data}-\text{model}$): (top)  maximum, (middle) average, and (bottom) RMS.}
    \begin{tabularx}{\linewidth}{|l|X|X|X|X|X|} 
        \hhline{| ------ |}
        \rule{0pt}{2.0ex}
        \makecell{Isotope} & 
        \multicolumn{1}{c|}{\scriptsize G4NDL} & 
        \multicolumn{1}{c|}{\scriptsize G4PhotoEvap.} & 
        \multicolumn{1}{c|}{\scriptsize NuDEX}  & 
        {\scriptsize G4CASCADE} & 
        \multicolumn{1}{c|}{\scriptsize G4CASCADE~(PE)} \\
        \hhline{| ------ |}
        \rule{0pt}{4.0ex}
        \makecell{ \ce{^12C}} \makecell{ Max \\ Mean \\ RMS} & \cellcolor{myRed} \makecell{27  \\   0.55 \\ 3.8} & \cellcolor{myRed!50!myPurple} \makecell{10 \\ -0.24 \\ 1.5} & \cellcolor{myBlue}\makecell{ 2.3 \\ -0.038 \\ 0.031}& \cellcolor{myBlue!50!myPurple}\makecell{ 0.20 \\ 0.0072 \\ 0.035} & \multicolumn{1}{p{2.6cm}|}{\cellcolor{myBlue!50!myPurple} \makecell*{0.20 \\  0.0072 \\  0.035}} \\
        \hhline{| ------ |}
        \rule{0pt}{4.0ex}
        \makecell{ \ce{^28Si}} \makecell{ Max \\ Mean \\ RMS}  & \cellcolor{myRed!50!myPurple} \makecell{  64 \\    -0.68  \\6.4} & \cellcolor{myRed} \makecell{100. \\ -1.5 \\ 20.} & \cellcolor{myBlue} \makecell{ 10. \\ -0.026 \\   1.4} & \cellcolor{myBlue} \makecell{ 10. \\ -0.028 \\   1.4} &  \multicolumn{1}{p{2.6cm}|}{\cellcolor{myBlue} \makecell{ 10. \\   -0.028 \\ 1.4}} \\
        \hhline{| ------ |}
        \rule{0pt}{4.0ex}
        \makecell{\ce{^40Ar}} \makecell{ Max \\ Mean \\ RMS}  & \cellcolor{myRed} \makecell{ 69 \\  0.48 \\  9.6} & \cellcolor{myRed!50!myPurple}   \makecell{ 58 \\  -1.6 \\ 8.5} &  \cellcolor{myPurple}  \makecell{13 \\ -0.74 \\ 2.0} &  \cellcolor{myBlue!50!myPurple}  \makecell{10 \\ -0.49 \\ 1.7} & 
        \multicolumn{1}{p{2.6cm}|}{\cellcolor{myBlue}   \makecell{10 \\ -0.49 \\ 1.6}} \\
        \hhline{| ------ |}
        \rule{0pt}{4.0ex}
        \makecell{ \ce{^48Ti}} \makecell{ Max \\ Mean \\ RMS}  & \cellcolor{myRed} \makecell{ 70. \\   -4.1 \\ 14} & \cellcolor{myRed!50!myPurple} \makecell{ 46 \\-4.6 \\ 11} & \cellcolor{myBlue} \makecell{8.7 \\ -0.089 \\  0.90 } & \cellcolor{myBlue!50!myPurple} \makecell{9.8 \\ -0.069 \\  1.0 } & 
        \multicolumn{1}{p{2.6cm}|}{\cellcolor{myPurple} \makecell{10. \\  -0.11 \\  1.1}} \\
        \hhline{| ------ |}
        \rule{0pt}{4.0ex}
        \makecell{ \ce{^56Fe}} \makecell{ Max \\ Mean \\ RMS}  & \cellcolor{myRed!50!myPurple} \makecell{ 49 \\ 0.48 \\ 5.4 }& \cellcolor{myPurple} \makecell{24 \\ 1.0 \\ 4.9 } &  \cellcolor{myRed} \makecell{  87 \\ 1.2 \\ 11} &  \cellcolor{myBlue} \makecell{  33 \\ -0.26 \\ 3.4} &  \multicolumn{1}{p{2.6cm}|}{\cellcolor{myBlue} \makecell{  36 \\  -0.27 \\  3.4}} \\
        \hhline{| ------ |}
        \rule{0pt}{4.0ex}
        \ce{^129Xe} \makecell{ Max \\ Mean \\ RMS}  & \cellcolor{myBlue!50!myPurple} \makecell{ 39  \\   -2.3 \\ 5.6 }& \cellcolor{myRed} \makecell{ 99 \\ -3.3 \\ 16 } &  \cellcolor{myBlue} \makecell{8.5 \\ -2.3 \\ 3.6} &  \cellcolor{myPurple} \makecell{47 \\ -0.63 \\ 7.1} &  \multicolumn{1}{p{2.6cm}|}{\cellcolor{myRed!50!myPurple} \makecell{ 38 \\  -2.7  \\ 7.4}} \\
        \hhline{| ------ |}
        \rule{0pt}{4.0ex}
        \ce{^155Gd} \makecell{ Max \\ Mean \\ RMS}  & \cellcolor{myBlue} \makecell{ 56  \\   1.3 \\ 11} & \cellcolor{myRed} \makecell{91 \\ 2.7 \\ 19 } &  \cellcolor{myBlue!50!myPurple} \makecell{81 \\ 3.2 \\ 14 } &  \cellcolor{myPurple} \makecell{91 \\ 3.1 \\ 17 } &  \multicolumn{1}{p{2.6cm}|}{\cellcolor{myRed!50!myPurple} \makecell{ 91 \\ 2.0 \\  18}} \\
        \hhline{| ------ |}
    \end{tabularx}
    \label{Table:residuals}
\end{table*}

Table~\ref{Table:residuals} shows summary comparisons for each model, measured by (top) maximum residuals compared to \CapGam, (middle) average residuals, and (bottom) root mean square (RMS) residuals. The residual is defined as the experimental (\CapGam) intensity minus the simulated intensity of a model.

For isotopes with mid-range atomic number $Z$, the default \CASCADE\ settings usually agree best with \CapGam. 
Unplaced \grs\ for high-$Z$ isotopes means that \GFNDL, which uses raw \gr\ intensities agnostic to level structure, sometimes performs better than \CASCADE.
However, it does not model correlations, and summation peaks from \GFNDL\ may be unreliable, as it often does not conserve energy.
In \CASCADE\ and completing de-excitations from unplaced \grs\ using \GFPE\ only sometimes improves agreement.
For low-$Z$ isotopes, both settings perform the same, since there are few unplaced \grs.

\section{Discussion}
\label{Sec:discussion}
\CASCADE's data-driven approach generally improves upon \GFPE\ and \GFNDL\ by agreeing better with data and conserving energy, which \GFNDL\ does not do, and it adds atomic de-excitation cascades following conversion electron emissions.
Intensities of \gr\ are often more accurately reproduced by \CASCADE\ than \GFPE: \CASCADE\ decreases the RMS residual by more than \SI{90}{\percent} for \ce{^12C}, \ce{^28Si}, and \ce{^{48}Ti}, by \SI{80}{\percent} for \ce{^40Ar}, and \SI{56}{\percent} for \ce{^{129}Xe}. 
Isotopes with many unplaced \gr\ lines tend to perform less well in \CASCADE.
\GFNDL\ often includes these lines, with occasional lines that do not appear in \ENSDF\ or have different intensity, but it does not properly correlate them or conserve energy.

For \ce{^56Fe}, \CapGam\ does not include the relative intensity of a \SI{14}{keV} \gr\ emitted by the decay of the first excited state of \ce{^57Fe}, likely due to the very high internal conversion coefficient for this transition.
Ref.~\cite{firestoneThermalNeutronCapture2017} reports recent measurement of the \ce{^56Fe}\ngamma\ce{^57Fe} reaction, including an updated level structure with the \SI{14}{\keV} transition and other additions.
\CASCADE\ uses this level structure instead of \ENSDF\ for neutron captures on \ce{^56Fe}.


Missing nuclear structure data can bias \gr\ intensities predicted by \CASCADE. 
Energy levels for which transitions to or from them are unknown are removed from \CASCADE's level structure.
If doing so also removes a transition with a large branching ratio, the branching ratio of other transitions from the same level are effectively enhanced, inflating the relative intensities of downstream \grs.
This effect is particularly pronounced in \ce{^129Xe} and \ce{^155Gd} in Fig.~\ref{fig:xenon_cascade}, where several incomplete energy levels cause significant bias in subsequent transitions.

The \NuDEX\ code~\cite{nudex} provides a similar but alternative model, not currently integrated into \Geant, which bases \grs\ on \ENSDF, with missing energy levels completed by the \DICEBOX\ statistical nuclear model code and \RIPL\ nuclear structure data~\cite{capoteRIPLReferenceInput2009}.
While \NuDEX\ and \CASCADE\ are data-driven and therefore fill similar niches, \CASCADE\ provides options to integrate unplaced \grs\ into the de-excitation cascade using \GFPE, comes fully-integrated into \Geant, and simulates atomic relaxation \xrs\ and Auger electrons following emission of an inner shell electron.
\NuDEX\ and \CASCADE\ reproduce \gr\ lines in \CapGam\ comparably well, though \NuDEX's use of the \DICEBOX\ statistical model is likely more accurate than \CASCADE's use of \GFPE. 
As a result, \NuDEX\ reproduces \CapGam's \gr\ lines for high-$Z$ isotopes more accurately than \CASCADE, since these isotopes typically have more unplaced \gr\ lines.

For \ce{^129Xe} and \ce{^155Gd} in particular, there are large gaps in the \SIrange{3}{4}{\MeV} range where no \gr\ lines are observed in experimental \gr\ intensity records.
For \ce{^129Xe}, Ref.~\cite{groshev1971spectra} measured only the primary \gr\ emissions, at the highest energies. Ref.~\cite{hamadaGammagammaDirectionalCorrelation1988a} measured correlations of lower-energy \grs\ to infer the structure of low-lying levels; however most secondary transitions from the daughter \ce{^130Xe} nucleus are not yet understood, explaining the gap observed around \SIrange{3}{4}{\MeV}. 
Ref.~\cite{massarczykNuclearDeformationNeutron2014} reports measured photon strength functions for other xenon nuclei, showing a significant statistical regime missing from \ENSDF\ data.
Likewise, measurements of \ce{^155Gd} in Refs.~\cite{kloraNuclearStructure156Gd1993,backlinLevels156GdStudied1982}
are used to infer energy levels up to \SI{3.2}{\MeV} as well as higher-energy \grs\ emitted in transitions from the neutron separation energy to low-lying states, leaving a gap around \SI{4}{\MeV} reflected in \ENSDF\ data, which photon strength function measurements reveal should be populated~\cite{baramsaiPhotonStrengthFunctions2013}. 
\NuDEX\ typically fills these gaps using a statistical model, indicating that there may be real \gr\ lines in these energy windows that \CASCADE\ does not produce, although more data are needed to experimentally verify them.
The gap at \SI{3.5}{MeV} for \ce{^129Xe} was partially filled in using data from a source external to \CapGam\ \cite{gelletlyXenonSpectrum1974}, but the gap at \SI{4}{MeV} for \ce{^155Gd} could not be filled in.

\section{Conclusions}
\CASCADE\ is a data-driven module for simulating \ngamma\ signals in \Geant. 
By addressing shortfalls in existing \ngamma\ in \Geant, \CASCADE\ will help improve background and signal acceptance calculations in MeV-scale rare event searches. 
The GitHub repository hosting \CASCADE\ includes code for generating the input binary files, so that new nuclear structure data or models can be integrated into the model as desired.
As such, additional level structure data for high-$Z$ nuclei would further improve \CASCADE's performance, as would measurements informing the structure of unbound levels, which may be particularly important for fast neutron-capture signals that become important in the \SIrange{11}{15}{\MeV} range. \CASCADE\ currently does not accurately simulate gamma emission timing, and releases all gammas for one event at the same time following capture. An area of future work is to improve this, likely using data-driven methods.
\CASCADE\ was developed for \Geant\texttt{-10}; an extension to \Geant\texttt{-11} is foreseen.
Additional updates are also planned to allow users to supplement level structures using \DICEBOX.

\section{Acknowledgements}
We would like to thank Luis Sarmiento Pico for help using {\tt G4RDAtomicDeexcitation} for atomic relaxation simulations.
We are also grateful to Ryan Krismer, Andrew Erlandson, and the \DEAP\ Collaboration for providing early use cases and invaluable beta testing of the \CASCADE\ code and documentation to ensure that it is ready-to-use and user-friendly prior to making it public.
This report is based upon work supported by the U.S. National Science Foundation (NSF) (Grants No. PHY-2310091 and PHY-2244610).

\bibliographystyle{deap} 
\bibliography{refs}

\end{document}